\documentclass[10pt]{article}
\usepackage{amssymb,amsmath,amsthm}
\usepackage{epsfig}
\newcommand{\ie}{\emph{i.e.}}
\newcommand{\cf}{\emph{cf}}

\newcommand{\demi}{\frac{1}{2}}
\newcommand{\Real}{\mathbb{R}}
\newcommand{\Nat}{\mathbb{N}}
\newcommand{\Sphere}{S}
\newcommand{\Smooth}{C}
\newcommand{\si}{L^1}
\newcommand{\sii}{L^2}
\newcommand{\sobi}{\mathop{W_0^{1,2}}\nolimits}
\newcommand{\Sobi}{\mathop{W^{1,2}}\nolimits}
\newcommand{\Dom}{\mathop{\mathrm{Dom}}\nolimits}
\newcommand{\supp}{\mathop{\mathrm{supp}}\nolimits}

\newcommand{\layer}{\mathop{\mathcal{L}}\nolimits}
\newcommand{\TotK}{\mathop{\mathcal{K}}\nolimits}
\newcommand{\TotM}{\mathop{\mathcal{M}}\nolimits}

\newcommand{\Hi}{\langle\mathsf{H}\mathrm{1}\rangle}
\newcommand{\Hii}{\langle\mathsf{H}\mathrm{2}\rangle}
\newenvironment{Assumption}[1]
{\begin{description}\item[$#1$]\quad}
{\end{description}}
\newtheorem{Theorem}{Theorem}
\newtheorem{Lemma}{Lemma}

\newtheorem{Corollary}{Corollary}
\theoremstyle{definition}
\newtheorem{Remark}{Remark}

\newtheorem{ex}{Example}
\newenvironment{Example}{\begin{ex}}{\qed\end{ex}}
\begin{document}
%
%
\noindent
{\LARGE\textbf{Topologically non-trivial quantum layers}}
\begin{quote}
{\large G.~Carron}
\footnote{
Electronic mail: Gilles.Carron@math.univ-nantes.fr
}
\\
\emph{\small
D\'epartement de Math\'ematiques, Universit\'e de Nantes, \\
2 rue de la Houssini\`ere,  BP 92208,  44\,322 Nantes cedex 03, France
}
\bigskip \\
{\large P.~Exner}
\footnote{
Also at \emph{Doppler Institute, Czech Technical University,
B\v{r}ehov{\'a}~7, 11519 Prague, Czech Republic};
Electronic mail: exner@ujf.cas.cz
}
\\
\emph{\small
Department of Theoretical Physics, Nuclear Physics Institute, \\
Academy of Sciences, 25068 \v{R}e\v{z} near Prague, Czech Republic
}
\bigskip \\
{\large D.~Krej\v{c}i\v{r}\'{\i}k}
\footnote{
On leave of absence from
\emph{Nuclear Physics Institute, Academy of Sciences,
25068 \v{R}e\v{z} near Prague, Czech Republic};
Electronic mail: dkrej@math.ist.utl.pt
}
\\
\emph{\small
Departamento de Matem\'atica, Instituto Superior T\'ecnico, \\
Av. Rovisco Pais, 1049-001 Lisboa, Portugal
}
\bigskip \\
{\small
Given a complete non-compact surface~$\Sigma$ embedded
in~$\Real^3$, we consider the Dirichlet Laplacian in the layer~$\Omega$
that is defined as a tubular neighbourhood  
of constant width about~$\Sigma$.
Using an intrinsic approach to the geometry of~$\Omega$,
we generalise the spectral results
of the original paper~\cite{DEK2} by Duclos \emph{et al.}
to the situation when~$\Sigma$ does not possess poles.
This enables us to consider topologically more complicated layers
and state new spectral results.
In particular, we are interested in layers built over surfaces with handles
or several cylindrically symmetric ends.
We also discuss more general regions obtained by
compact deformations of certain~$\Omega$.
}
\end{quote}
%
%
\newpage
%
%
\section{Introduction}
The spectral properties of the Dirichlet Laplacian in infinitely
stretched regions have attracted a lot of attention
since the existence of geometrically induced discrete spectrum
for certain strips in the plane was proved in~\cite{ES}.
The study was motivated by mesoscopic physics where a reasonable
model for the dynamics of a particle in quantum waveguides
is given by the Laplacian in hard-wall tubular neighbourhoods
of infinite curves in~$\Real^d$, $d=2,3$ (quantum strips, tubes),
or surfaces in~$\Real^3$ (quantum layers);
see~\cite{DE,LCM} for the physical background and references.
Nowadays, it is well known that any non-trivial curvature
of the reference curve, that is asymptotically straight,
produces bound states below the essential spectrum
in the strips and tubes,~\cite{DE,GJ,RB}.

\begin{figure}[h]
\begin{center}
\epsfig{file=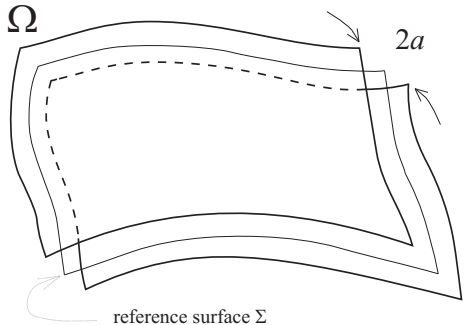,width=0.5\textwidth}
\end{center}
\caption{
The configuration space~$\Omega$ defined by~(\ref{tube})
as the space delimited by two parallel surfaces
at the distance~$a$ from~$\Sigma$.
}
\label{Fig.layer}
\end{figure}

The analogous problem in curved layers is much more complicated
and it was investigated quite recently in~\cite{DEK2,DEK1,EK3}.
Let~$\Sigma$ be a complete non-compact surface embedded in~$\Real^3$,
$\Omega$~be a tube of radius~$a>0$ about~$\Sigma$,
\ie\/ (see Figure~\ref{Fig.layer}),
\begin{equation}\label{tube}
  \Omega:=\left\{z\in\Real^3 \ | \ \textrm{distance}\,(z,\Sigma)<a\right\},
\end{equation}
and~$-\Delta_D^\Omega$ denote the Dirichlet Laplacian in~$\sii(\Omega)$.
If the surface is a locally deformed plane, the existence
of bound states below the essential spectrum of the Laplacian
was demonstrated in~\cite{DEK1}.
A more general situation was treated in~\cite{DEK2};
assuming that~$\Sigma$ is non-trivially curved, it has asymptotically
vanishing curvatures and possesses a pole,
we found several sufficient conditions
which guarantee the existence of discrete spectrum.
Finally, let us mention that an asymptotic expansion
of the ground-state eigenvalue in layers built over
mildly curved planes was found in~\cite{EK3}.
%

\begin{figure}[h]
\begin{center}
\epsfig{file=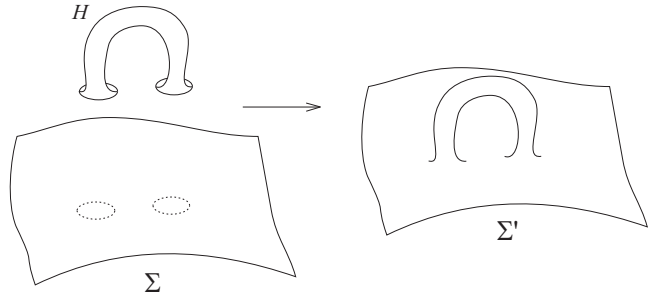,width=0.7\textwidth}
\end{center}
\caption{
Surface with a handle~$\Sigma'$ is constructed from~$\Sigma$
by attaching smoothly to it a curved cylindrical surface~$H$.
By virtue of~Corollary~\ref{Cor.CV},
one handle is sufficient to achieve the condition~(a)
of Theorem~\ref{thm.main}.
}
\label{Fig.handle}
\end{figure}

While the paper~\cite{DEK2} covers a wide class of layers,
the technical requirement about the existence of a pole
on~$\Sigma$ (\ie, the exponential map is a diffeomorphism)
restricted substantially the topological structure of the reference surface.
In particular, $\Sigma$~was necessarily diffeomorphic to~$\Real^2$
and as such it was simply connected.
The main goal of this paper is to extend the sufficient
conditions established in~\cite{DEK2}
without assuming the existence of poles on~$\Sigma$ and
without making any other (unnatural) topological and geometrical assumptions.
In addition to this substantial generalisation,
we will derive particularly interesting spectral
results for quantum layers built over surfaces with handles
or several cylindrically symmetric ends
(see Figure~\ref{Fig.handle} and~\ref{Fig.ends}).

Let us recall the reason why the existence of a pole on~$\Sigma$
was required in~\cite{DEK2}.
According to the usual strategy used in the spectral theory
of quantum waveguides,
one expresses the Laplacian~$-\Delta_D^\Omega$
in the pair of coordinates~$(x,u)$, where~$x$ parameterises
the reference surface~$\Sigma$ and~$u\in(-a,a)$ its normal bundle.
Assuming the existence of a pole,
$\Sigma$~could be parameterised globally by means of geodesic polar coordinates,
which were well suited for the construction of explicit mollifiers on~$\Sigma$
needed to regularise generalised trial functions establishing
the existence of bound states below the essential spectrum.

There are several possibilities how to treat surfaces without poles.
Since the above mentioned regularisation is needed
out of a compact part of~$\Sigma$ only,
one way is to replace the polar coordinates by geodesic coordinates
based on a curve enclosing the interior part.
This approach is well suited for surfaces of one end
(see the definition below),
however, it has to be modified in more general situations.
In this paper, we introduce a different strategy which does not require
any special choice of coordinates on~$\Sigma$.
We employ substantially a consequence of~\cite{Huber}
that if the Gauss curvature is integrable then there always exists
a sequence of functions on~$\Sigma$ having the properties
of the mollifiers mentioned above.

\section{Statement of Results}
\begin{figure}[h]
\begin{center}
\epsfig{file=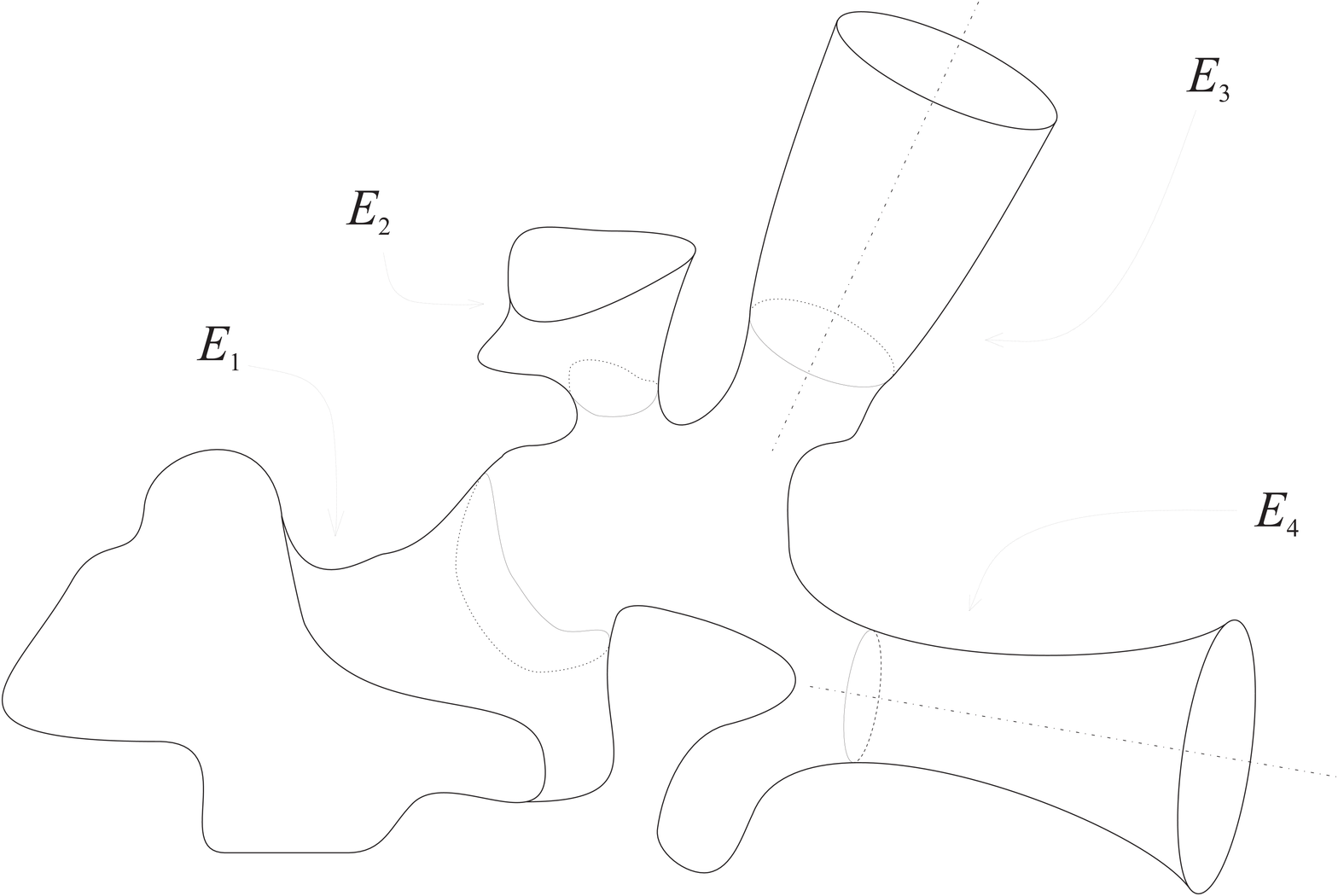,width=0.8\textwidth}
\end{center}
\caption{
Surface with four ends ($E_1, \dots, E_4$).
By virtue of Theorem~\ref{thm.ends},
each cylindrically symmetric end ($E_3,E_4$)
with a positive total Gauss curvature
and curvatures vanishing at infinity
produces at least one discrete eigenvalue.
}
\label{Fig.ends}
\end{figure}

To state here the main results we need to introduce some notation
and basic assumptions.
Let~$\kappa_1^2$ denote the spectral threshold of the planar layer
of width~$2a$, \ie~$\kappa_1:=\pi/(2a)$.
The induced metric on~$\Sigma$ and the corresponding covariant
derivative will be denoted by~$g$ and~$\nabla_{\!g}$, respectively.
Let~$K$,~$M$ and~$k_\pm$ denote, respectively,
the Gauss curvature, the mean curvature and the principal curvatures of~$\Sigma$.
Denoting by~$d\Sigma$ the surface area-element, we may define
the total Gauss curvature~$\TotK$ and the total mean curvature~$\TotM$,
respectively, by the integrals
\begin{equation}\label{Tot}
  \TotK:=\int_\Sigma K\,d\Sigma
  \qquad\textrm{and}\qquad
  \TotM^2:=\int_\Sigma M^2\,d\Sigma.
\end{equation}
The latter always exists (it may be~$+\infty$), while the former
is well defined provided
\begin{Assumption}{\Hi}
  $K \in \si(\Sigma)$,
\end{Assumption}
which will be a characteristic assumption of this work.
Henceforth, we shall also assume that~$k_\pm$ are bounded and
\begin{Assumption}{\Hii}
  $a < \rho_m:=\left(\max\{\|k_+\|_\infty,\|k_-\|_\infty\}\right)^{-1}$
  \quad and \quad
  $\Omega$ does not overlap,
\end{Assumption}
which we need in order to ensure that the layer~$\Omega$
is a submanifold of~$\Real^3$.
An open set~$E\subseteq\Sigma$
is called an \emph{end} of~$\Sigma$ if it is connected, unbounded
and if its boundary~$\partial E$ is compact (see Figure~\ref{Fig.ends});
its total curvatures are defined by means of~(\ref{Tot})
with the domain of integration being the subset~$E$ only.
We say that a manifold embedded in~$\Real^3$ is cylindrically symmetric
if it is invariant under rotations about a fixed axis in~$\Real^3$.
Our main result reads as follows.
\begin{Theorem}\label{thm.main}
Let~$\Sigma$ be a complete non-compact connected orientable surface
of class~$\Smooth^2$ embedded in~$\Real^3$ and satisfying~$\Hi$.
Let the layer~$\Omega$ defined by~\emph{(\ref{tube})}
as the tube of radius~$a>0$ about~$\Sigma$ satisfy~$\Hii$.
\begin{itemize}
\item[\emph{(i)}]
If the curvatures~$K$ and~$M$ vanish at infinity of~$\Sigma$,
then
$$
  \inf\sigma_\mathrm{ess}\left( -\Delta_D^\Omega \right) = \kappa_1^2.
$$
\item[\emph{(ii)}]
If the surface~$\Sigma$ is not a plane, then any of the conditions
\begin{itemize}
\item[\emph{(a)}]
$\TotK \leq 0$
\vspace{-3.5ex} \\
\item[\emph{(b)}]
$a$ is small enough
\ and \ $\nabla_{\!g} M \in \sii_\mathrm{loc}(\Sigma)$
\vspace{-3.5ex} \\
\item[\emph{(c)}]
$\TotM=+\infty$
\ and \ $\nabla_{\!g} M \in \sii(\Sigma)$
\vspace{-3.5ex} \\
\item[\emph{(d)}]
$\Sigma$ contains a cylindrically symmetric end
with a positive total Gauss curvature
\end{itemize}
is sufficient to guarantee that
$$
  \inf\sigma\left( -\Delta_D^\Omega \right) < \kappa_1^2.
$$
\end{itemize}
Consequently, if the surface~$\Sigma$ is not a plane
but its curvatures vanish at infinity,
then any of the conditions \emph{(a)--(d)}
is sufficient to guarantee that~$-\Delta_D^\Omega$ has at least
one eigenvalue of finite multiplicity below 
the threshold of its essential spectrum,
\ie~$\sigma_\mathrm{disc}\left( -\Delta_D^\Omega \right) \not= \emptyset$.
\end{Theorem}

Let us compare this theorem with the results obtained in~\cite{DEK2}.
An improvement concerns the essential spectrum.
While only a lower bound on the threshold was found in~\cite{DEK2},
here we shall use known results about the spectral threshold
of complete surfaces in order to show that the essential spectrum
starts just at~$\kappa_1^2$.
The conditions (a)--(d) are adopted from~\cite{DEK2},
however, we do not assume that~$\Sigma$ is of class~$\Smooth^3$
in~(b) and~(c) of Theorem~\ref{thm.main},
which was required in~\cite{DEK2} in order to give
a meaning to~$\nabla_{\!g} M$. Indeed, only the integrability conditions
on the gradient are needed.

The most significant generalisation concerning
all the results is that we have got rid of the strong
assumption about the existence of a pole on~$\Sigma$.
Actually, Theorem~\ref{thm.main} involves quantum layers built over
general surfaces without any additional hypotheses about the existence
of a special global parameterisation, the number of ends, and other topological
and geometrical restrictions.

An interesting new spectral result then follows from the observation
that making the topology of~$\Sigma$ more complicated
than that of the plane, one always achieves that 
the basic condition~(a) is satisfied.
\begin{Corollary}\label{Cor.CV}
Under the assumptions of \emph{Theorem~\ref{thm.main}},
one has
$
  \inf\sigma(-\Delta_D^\Omega)<\kappa_1^2
$
whenever~$\Sigma$ is not conformally equivalent to the plane.
\end{Corollary}
\noindent
Indeed, the Cohn-Vossen inequality~\cite{Cohn-Vossen} yields
\begin{equation}\label{Gauss-Bonnet}
  \TotK \leq 2\pi\,(2-2h-e),
\end{equation}
where~$h$ is the genus of~$\Sigma$, \ie~the number of handles,
and~$e$ is the number of ends.
In particular, the condition~(a) of Theorem~\ref{thm.main}
is always fulfilled whenever the surface is not simply connected.
%
\begin{Example}\label{Ex.paraboloid}
Let~$\Sigma$ be the elliptic paraboloid.
It is easy to check that it has curvatures vanishing at infinity
and that the condition~(c) of Theorem~\ref{thm.main} is always fulfilled.
On the other hand, it violates the condition~(d)
whenever it is not a paraboloid of revolution,
and the condition~(a) does not hold because
the total Gauss curvature is always equal to~$2\pi$.
Attaching a handle to~$\Sigma$, the total curvature becomes
equal to~$-2\pi$.
%
\begin{figure}[h]
\begin{center}
\epsfig{file=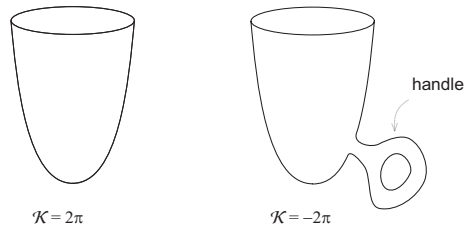,height=0.15\textheight}
\end{center}
\caption{
Elliptic paraboloid (without or with one handle attached, respectively)
of Example~\ref{Ex.paraboloid}.
}
\end{figure}
\end{Example}

It was proved in~\cite{DEK2} that any layer built over
a cylindrically symmetric surface diffeomorphic to~$\Real^2$
had a spectrum below the energy~$\kappa_1^2$.
Since this class of reference surfaces may have a non-negative
total Gauss curvature only, it gave an important alternative condition to~(a)
in the case~$\TotK>0$.
In Theorem~\ref{thm.main}, an interesting generalisation
to~\cite{DEK2} is introduced by virtue of the condition~(d),
where it is supposed now that only an unbounded subset of~$\Sigma$
admits a cylindrical symmetry at infinity (see Figure~\ref{Fig.ends}).
This extension is possible due to the fact that
the sequence of trial functions
establishing the existence of spectrum below~$\kappa_1^2$
for surfaces of revolution with~$\TotK>0$
was ``localised at infinity''
(\ie, for any compact set of~$\Omega$, there is an element
from the sequence supported out of the compact).
Consequently, it may be localised just at the end satisfying
the condition~(d) of Theorem~\ref{thm.main}.
Since any deformation of a bounded part of~$\Omega$ does not
affect this spectral result, we may consider more general
regions than tubes~(\ref{tube}).
What is important is that such local deformations
do not include only bends and protrusions which are traditionally
a source of binding, but constrictions as well.
Moreover, since such trial functions localised at different ends
will be orthogonal as elements of~$\sii(\Omega)$,
we may produce an arbitrary number of bound states
by attaching to~$\Omega$ a sufficient number of suitable outlets.
Finally, since the essential spectrum is stable under compact deformation
of~$\Omega$, we arrive at the following result.
%
\begin{Theorem}\label{thm.ends}
Let~$\Omega$ be a layer~\emph{(\ref{tube})} satisfying~$\Hi$, $\Hii$
and the condition~\emph{(i)} of \emph{Theorem~\ref{thm.main}}.
Assume that the reference surface~$\Sigma$ contains~$N \geq 1$
cylindrically symmetric ends, each of them having
a positive total Gauss curvature.
Let~$\Omega'$ be an unbounded region without boundary in~$\Real^3$
obtained by any compact deformation of~$\Omega$.
Then
\begin{itemize}
\item[\emph{(i)}]
$\inf\sigma_\mathrm{ess}\big( -\Delta_D^{{\Omega'}} \big) = \kappa_1^2$\,,
\item[\emph{(ii)}]
there will be at least~$N$ eigenvalues in~$\left(0,\kappa_1^2\right)$,
with the multiplicity taken into account.
\end{itemize}
\end{Theorem}
\begin{Example}\label{Ex.cone}
Fix $\theta\in(0,\frac{\pi}{2})$ and consider
the conical region~$\Omega'$ in~$\Real^3$
given by rotating the planar region (see Figure~\ref{Fig.cone}):
\begin{equation*}
  \left\{
  (x,y)\in\Real^2 \,|\ (x,y)\in
  \big((0,2a\cot\theta)\times(0,x\tan\theta]\big)
  \cup
  \big([2a\cot\theta,\infty)\times(0,2a)\big)
  \right\}
\end{equation*}
along the axis $y=x\tan\theta$ in~$\Real^3$.
Note that~$\Omega'$ is not a layer~(\ref{tube}) because of the singularity
of the conical surface. Nevertheless, it may be considered as a compact
deformation of the layer built over a smoothed cone
whose total Gauss curvature is equal to~$2\pi(1-\sin\theta) \in (0,2\pi)$.
Consequently, we know that~$-\Delta_D^{\Omega'}$ possesses
at least one discrete eigenvalue below~$\kappa_1^2$
due to Theorem~\ref{thm.ends}.
This is a non-trivial result for flat enough conical layers only,
since using a trick analogous to that of~\cite{ABGM}
one can check that the cardinality 
of~$\sigma_\mathrm{disc}(-\Delta_D^{\Omega'})$
can exceed any fixed integer for~$\theta$ small enough.
%
\begin{figure}[h]
\begin{center}
\epsfig{file=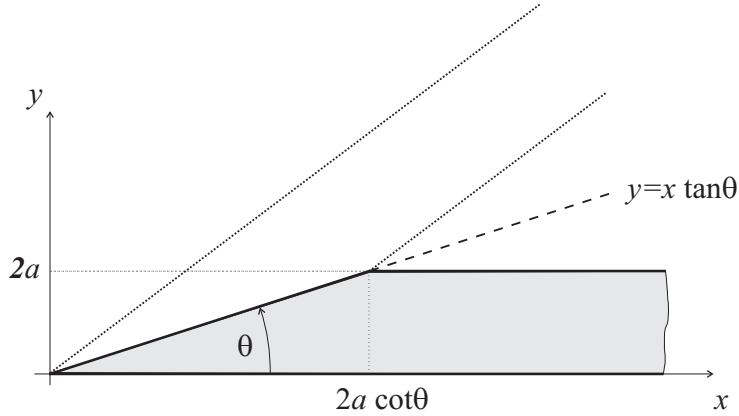,width=0.8\textwidth}
\end{center}
\caption{
The planar region of Example~\ref{Ex.cone}.
}
\label{Fig.cone}
\end{figure}
\end{Example}
%

\section{Preliminaries}
Let~$\Sigma$ be a connected orientable surface of class~$\Smooth^2$
embedded in~$\Real^3$.
The orientation can be specified by the choice of a globally
defined unit normal vector field, $n:\Sigma\to\Sphere^2$,
which is a function of class~$\Smooth^1$.
For any~$x\in\Sigma$, the Weingarten map
\begin{equation}\label{Weingarten}
  L_x : \ T_x\Sigma \to T_x\Sigma : \,
  \left\{ \xi \mapsto -dn_x(\xi) \right\}
\end{equation}
defines the principal curvatures~$k_\pm$ of~$\Sigma$
as its eigenvalues with respect to the induced metric~$g$.
The Gauss curvature and the mean curvature are defined
by~$K:=k_+ k_-$ and $M:=\demi(k_+ + k_-)$, respectively,
and are continuous functions on~$\Sigma$.

Put~$a>0$. We define a layer~$\Omega$ of width~$2a$
as the image of the mapping
\begin{equation}\label{layer}
  \layer : \ \Sigma\times(-a,a) \to \Real^3 : \,
  \left\{(x,u) \mapsto x+u\,n(x) \right\}.
\end{equation}
Henceforth, we shall always assume~$\Hii$.
Then~$\layer$ induces a diffeomorphism and~$\Omega$ is a submanifold
of~$\Real^3$ corresponding to the set of points squeezed
between two parallel surfaces at the distance~$a$ from~$\Sigma$
(see Figure~\ref{Fig.layer}),
\ie, if~$\Sigma$ does not have a boundary then the definition of~$\Omega$
via~(\ref{layer}) and~(\ref{tube}) are equivalent.
We shall identify it with the Riemannian manifold $\Sigma\times(-a,a)$
endowed with the metric~$G$ induced by the immersion~(\ref{layer}).
One has
\begin{equation}\label{metric}
  G=g\circ\left(I_x-u\,L_x\right)^2 + du^2
  \qquad\textrm{and}\qquad
  d\Omega=\left(1-2Mu+Ku^2\right) d\Sigma\,du,
\end{equation}
where~$I_x$ denotes the identity map on~$T_x\Sigma$
and~$d\Omega$ stands for the volume element of~$\Omega$.
It is worth to notice that~(\ref{metric}) together with~$\Hii$
yields that~$G$ can be estimated by the surface metric,
\begin{equation}\label{g<G<g}
  C_- g + du^2 \leq G \leq C_+ g + du^2,
  \qquad\textrm{where}\quad\
  C_\pm:=\left(1 \pm a\rho_m^{-1}\right)^2.
\end{equation}
\begin{Remark}\label{Rem.Diff}
Formally, it is possible to consider $(\Sigma\times(-a,a),G)$
as an abstract Riemannian manifold where only
the surface~$\Sigma$ is embedded in~$\Real^3$.
Then we do not need to assume the second part of~$\Hii$,
\ie~``$\Omega$ does not overlap''.
\end{Remark}

We denote by~$-\Delta_D^\Omega$, or simply~$-\Delta$,
the Dirichlet Laplacian on~$\sii(\Omega)$.
We shall consider it in a generalised sense as the operator
associated with the Dirichlet form
\begin{equation}\label{form}
  Q(\psi,\phi) := \int_\Omega
  \left\langle\nabla\psi,\nabla\phi\right\rangle \,d\Omega
  \qquad\textrm{with}\qquad
  \Dom Q := \sobi(\Omega) ,
\end{equation}
Here~$\nabla$ is the gradient corresponding to the metric~$G$
and~$\langle\cdot,\cdot\rangle$ denotes the inner product
in the manifold~$\Omega$ induced by~$G$; 
the associated norm will be denoted by~$|\cdot|$.
Similarly, the inner product and the norm in the Hilbert space~$\sii(\Omega)$
will be denoted by~$(\cdot,\cdot)$ and~$\|\cdot\|$, respectively.
We shall sometimes abuse the notation slightly by writing
$(\cdot,\cdot) \equiv \int_\Omega \langle\cdot,\cdot\rangle d\Omega$
and $\|\cdot\| \equiv \int_\Omega |\cdot| d\Omega$ for vector fields.
The subscript~``$g$'' will be used in order to distinguish
similar objects associated to the surface~$\Sigma$.

Since the quadratic form~$Q$ is densely defined, symmetric, positive
and closed on its domain, the corresponding Laplacian~$-\Delta$
is a positive self-adjoint operator.
Denoting by~$(x^\mu) \equiv (x^1,x^2)$ local coordinates for~$\Sigma$
and by~$G^{ij}$ the coefficients of the inverse of~$G$
in the coordinates~$(x^i) \equiv (x^\mu,u)$ for~$\Omega$, we can write
\begin{equation}\label{Laplacian}
  -\Delta
  \ = \ -|G|^{-\demi}\partial_i |G|^\demi G^{ij} \partial_j
  \ = \ -|G|^{-\demi}\partial_\mu |G|^\demi G^{\mu\nu} \partial_\nu
  -\partial_u^2 + 2 M_u\,\partial_u 
\end{equation}
in the form sense, where~$|G|:=\det{G}$ and
\begin{equation}\label{mean}
  M_u := \,\frac{M-Ku}{1-2Mu+Ku^2} \,,
\end{equation}
which is the mean curvature of the parallel surface
$\layer(\Sigma\times\{u\})$.

The above definitions of~$\Omega$ and the corresponding
Dirichlet Laplacian are valid for any orientable surface~$\Sigma$
of class~$\Smooth^2$ provided~$\Hii$
(or its first part only in view of Remark~\ref{Rem.Diff}) holds true.
Nevertheless, since we are interested in the existence of discrete spectrum
of~$-\Delta_D^\Omega$, and it always exists whenever~$\Omega$
is bounded, in the sequel we shall assume that
\begin{center}
  $\Sigma$ is \emph{complete} and \emph{non-compact}.
\end{center}

It is easy to see that the spectrum of the planar layer
$\Omega_0:=\Real^2\times(-a,a)$ is purely continuous
and coincides with the interval~$[\kappa_1^2,\infty)$,
where the threshold is the first eigenvalue of the Dirichlet
Laplacian on the transverse section, \ie~\mbox{$\kappa_1:=\pi/(2a)$}.
In what follows we shall use the corresponding normalised eigenfunction
given explicitly by
\begin{equation}\label{trans.EF}
  \chi_1(u):=\sqrt{\frac{1}{a}}\,\cos{\kappa_1 u}.
\end{equation}
Using the identities~$|\nabla u|=1$ and~$-\Delta u=2M_u$, we get
\begin{equation}\label{Laplace.chi}
  -\Delta\chi_1(u)
  \,=\, 2 M_u\,\chi_1'(u)
  +\kappa_1^2\,\chi_1(u) \,.
\end{equation}
%

\section{Essential Spectrum}
We shall localise the essential spectrum of~$-\Delta_D^\Omega$
for \emph{asymptotically planar} layers,
\ie, the curvatures of~$\Sigma$ vanish at infinity
which we abbreviate by
\begin{equation}\label{planar}
  K,M \xrightarrow[]{\ \infty\ } 0.
\end{equation}
Recall that a function~$f$, defined on a non-compact manifold~$\Sigma$,
is said to vanish at infinity if
$$
  \forall\epsilon>0 \quad
  \exists R_\epsilon>0,\, x_\epsilon\in\Sigma \quad
  \forall x\in\Sigma\setminus \overline{B(x_\epsilon,R_\epsilon)} \,: \quad
  |f(x)|<\epsilon \,,
$$
where $B(x_\epsilon,R_\epsilon)$ denotes the open ball of centre~$x_\epsilon$
and radius~$R_\epsilon$.
The property~(\ref{planar}) is equivalent to the vanishing
of the principal curvatures, \ie~$k_\pm \xrightarrow[]{\infty} 0$.

The proof of statement (i) of Theorem~\ref{thm.main}
is achieved in two steps.
If the layer is asymptotically planar, then it was shown in~\cite{DEK2}
that the essential spectrum of~$-\Delta_D^\Omega$ is bounded
from below by~$\kappa_1^2$ provided the surface possesses a pole.
Here we adapt the proof (based on a Neumann bracketing argument)
to the case of any complete surface with asymptotically vanishing curvatures.
In the second part of this section,
we establish the opposite bound on the threshold
by means of a different method.

\subsection{Lower Bound,
$\inf\sigma_\mathrm{ess}(-\Delta_D^\Omega)\geq\kappa_1^2$}
Fix an~$\epsilon>0$ and consider a precompact region
$\mathcal{B} \supseteq B(x_\epsilon,R_\epsilon)$
with $\Smooth^1$-smooth boundary such that
\begin{equation}\label{estimates}
  \forall (x,u)\in\Omega_\mathrm{ext} \,: \
  (1-a\epsilon)^2
  \leq
  1-2M(x)\,u+K(x)\,u^2
  \leq
  (1+a\epsilon)^2,
\end{equation}
where $\Omega_\mathrm{ext}:=\Omega\setminus\overline{\Omega}_\mathrm{int}$
with $\Omega_\mathrm{int}:=\mathcal{B}\times(-a,a)$.
Denote by~$-\Delta_N$ the Laplacian $-\Delta_D^\Omega$
with a supplementary Neumann boundary condition
on~$\partial\mathcal{B}\times(-a,a)$, that is,
the operator associated with the form
$Q_N:=Q_N^\mathrm{int} \oplus Q_N^\mathrm{ext}$,
where
\begin{equation*}
  Q_N^\omega(\psi,\phi) := \int_{\Omega_\omega}
  \left\langle\nabla\psi,\nabla\phi\right\rangle \,d\Omega,
  \quad
  \Dom Q_N^\omega := \left\{
  \psi\in\Sobi(\Omega_\omega)\,|\
  \psi(\cdot,\pm a)=0
  \right\}
\end{equation*}
for $\omega\in\{\mathrm{int},\mathrm{ext}\}$.
Since $-\Delta_D^\Omega \geq -\Delta_N$
and the spectrum of the operator associated to~$Q_N^\mathrm{int}$
is purely discrete, \cf~\cite[Chap.~7]{Davies},
the minimax principle gives the estimate
$$
  \inf\sigma_\mathrm{ess}(-\Delta_D^\Omega)
  \geq
  \inf\sigma_\mathrm{ess}(-\Delta_N^\mathrm{ext})
  \geq
  \inf\sigma(-\Delta_N^\mathrm{ext}),
$$
where~$-\Delta_N^\mathrm{ext}$ denotes the operator
associated to~$Q_N^\mathrm{ext}$.
Neglecting the non-negative ``longitudinal'' part of the Laplacian
(\ie, the first term at the r.h.s. of~(\ref{Laplacian}))
and using the estimates~(\ref{estimates}),
we arrive easily at the following lower bound
$$
  -\Delta_N^\mathrm{ext}
  \geq
  \left(\frac{1-a\epsilon}{1+a\epsilon}\right)^2 \kappa_1^2
  \qquad\textrm{in}\quad
  \sii(\Omega_\mathrm{ext}),
$$
which holds in the form sense
(see also proof of Theorem~4.1 in~\cite{DEK2}).
The claim then follows by the fact that~$\epsilon$
can be chosen arbitrarily small.
\hfill\qed
%

\subsection{Upper Bound,
$\inf\sigma_\mathrm{ess}(-\Delta_D^\Omega)\leq\kappa_1^2$}
It follows from~\cite{Donnelly-81} that if $K \xrightarrow[]{\infty} 0$
then the threshold of the (essential) spectrum of the Laplacian on~$\Sigma$,
$-\Delta_g$, equals~$0$.
This is equivalent to the statement that
for any~$\varepsilon>0$ there exists an infinite-dimensional subspace
$\mathcal{D}_{\!g}\subseteq\Smooth_0^\infty(\Sigma)$
such that
\begin{equation}\label{ess.sp.surface}
  \forall\varphi\in\mathcal{D}_{\!g}: \quad
  \|\nabla_{\!g}\varphi\|_g \leq \varepsilon \|\varphi\|_g.
\end{equation}
It is easy to see that the following identity holds true
\begin{equation}\label{identity}
  \forall\varphi\in\Smooth_0^\infty(\Sigma): \quad
  \|\nabla\varphi\chi_1\|^2 = \||\nabla\varphi|\,\chi_1\|^2
  -\left(\varphi\chi_1,\varphi\Delta\chi_1\right).
\end{equation}
Using the estimates~(\ref{g<G<g}) and~(\ref{ess.sp.surface}),
we have
$$
  \||\nabla\varphi|\,\chi_1\|^2
  \leq
  (C_+/{C_-}^2) \, \varepsilon^2 \, \|\varphi\,\chi_1\|^2,
$$
while the second term at the r.h.s. of~(\ref{identity})
can be rewritten by means of~(\ref{Laplace.chi}) as follows
$$
  -\left(\varphi\Delta\chi_1,\varphi\chi_1\right)
  =\kappa_1^2\,\|\varphi\,\chi_1\|^2
  +(\varphi\chi_1',2M_u\varphi\chi_1).
$$
Integrating by parts w.r.t.~$u$ in the second term at the r.h.s.
of the last equality, we conclude from~(\ref{identity})
that for any~$\varepsilon>0$ there exists
$
  \mathcal{D}:=\mathcal{D}_{\!g}\otimes\{\chi_1\}\subset\Smooth_0^\infty(\Omega)
$
such that
$$
  \forall\psi\in\mathcal{D}: \quad
  \|\nabla\psi\|^2-(\psi,K_u\psi)
  \leq
  \left(\kappa_1^2+(C_+/{C_-}^2)\,\varepsilon^2\right) \|\psi\|^2,
$$
where
$$
  K_u := \,\frac{K}{1-2Mu+Ku^2} 
$$
is the Gauss curvature of the parallel surface~$\layer(\Sigma\times\{u\})$. 
This proves that $\inf\sigma_\mathrm{ess}(-\Delta-K_u)\leq\kappa_1^2$.
Since~$K_u$ vanishes at infinity by the assumption~(\ref{planar}),
\ie, the operator $K_u(-\Delta+1)^{-1}$ is compact in~$\sii(\Omega)$,
the same spectral result holds for the operator~$-\Delta$.
\hfill\qed
\begin{Remark}
Notice that only $K \xrightarrow[]{\infty} 0$ is needed in order
to establish the upper bound.
\end{Remark}
%

\section{Geometrically Induced Spectrum}
It was shown in the precedent section that the threshold of the essential
spectrum is stable under any deformation of the planar layer
such that the deformed layer is still planar asymptotically
in the sense of~(\ref{planar}).
The aim of this section is to prove the sufficient conditions (a)--(d)
of the second part of Theorem~\ref{thm.main}, which guarantee
the existence of spectrum below the energy~$\kappa_1^2$.
Since the spectral threshold of the planar layer is just~$\kappa_1^2$,
the spectrum below this value is induced by the curved geometry
and it consists of discrete eigenvalues if the layer is asymptotically planar.

All the proofs here are based on the variational idea of finding
a trial function~$\Psi$ from the form domain of~$-\Delta_D^\Omega$
such that
\begin{equation}\label{shifted.form}
  Q_1[\Psi] := Q[\Psi]-\kappa_1^2\,\|\Psi\|^2 \,<\, 0.
\end{equation}
The important technical tool needed to establish the conditions~(a)--(c)
is the existence of appropriate mollifiers on~$\Sigma$ which is
ensured by the following lemma.
\begin{Lemma}\label{Huber}
Assume~$\Hi$. Then there exists a sequence~$\{\varphi_n\}_{n\in\Nat}$
of smooth functions with compact supports in~$\Sigma$ such that
\begin{enumerate}
\item
$\forall n\in\Nat: \ 0 \leq \varphi_n \leq 1$,
\vspace{-3.5ex} \\
\item
$\|\nabla_{\!g}\varphi_n\|_g \xrightarrow[n\to\infty]{} 0$,
\vspace{-3.5ex} \\
\item
$\varphi_n \xrightarrow[n\to\infty]{} 1 \ $ uniformly on compacts of~$\Sigma$.
\end{enumerate}
\end{Lemma}
\begin{proof}
If~$\Hi$ holds true then it follows from~\cite{Huber}
that~$(\Sigma,g)$ is conformally equivalent
to a closed surface from which a finite number of points have been removed.
However, the integral $\|\nabla_{\!g}\varphi_n\|_g$
is a conformal invariant and it is easy to find a sequence
having the required properties on the ``pierced'' closed surface.
\end{proof}

\noindent
This sequence enables us to regularise a generalised trial function
which would give formally a negative value of the functional~(\ref{shifted.form}),
however, it is not integrable in~$\sii(\Sigma)$.
Since the trial functions used below are adopted from~\cite{DEK2}
and the proofs using different mollifiers of Lemma~\ref{Huber}
requires just slight modifications, we will not go into great details
in the proofs of conditions~(a)--(c).
The sufficient condition~(d) does not use the mollifiers of Lemma~\ref{Huber}.
This condition is established by means of the fact
that the sequence of trial functions
employed in~\cite{DEK2} for cylindrically symmetric layers
was localised only at infinity of the layer.

\subsection{
Condition (a)
}
Using the first transverse mode~(\ref{trans.EF})
as the generalised trial function, one gets
$$
  Q_1[\varphi_n\chi_1]=
  \left\| |\nabla\varphi_n|\,\chi_1\right\|^2
  +\left(\varphi_n,K\varphi_n\right)_g.
$$
Since~$|\nabla\varphi_n|$ can be estimated by~$|\nabla_{\!g}\varphi_n|_g$
by means of~(\ref{g<G<g}), the first term at the r.h.s. tends
to zero as~$n\to\infty$ due to Lemma~\ref{Huber}.
The second one tends to the total Gauss curvature~$\TotK$
because of Lemma~\ref{Huber} and the dominated convergence theorem.
Hence, if~$\TotK<0$, we can find a finite~$n_0$ such
that $Q_1[\varphi_{n_0}\chi_1]<0$.

In the critical case, \ie~$\TotK=0$, one adds to~$\varphi_n\chi_1$
a small deformation term. Let~$\varepsilon$ be a real number,
which will be specified later,
and let~$j$ be an infinitely smooth positive function on~$\Sigma$
with a compact support in a region where the mean curvature~$M$
is non-zero and does not change sign.
Defining \mbox{$\theta(x,u):=j(x)u\chi_1(u)$}, one can write
$$
  Q_1[\varphi_n\chi_1+\varepsilon\theta]
  = Q_1[\varphi_n\chi_1] + 2\varepsilon\, Q_1(\theta,\varphi_n\chi_1)
  +\varepsilon^2 Q_1[\theta].
$$
Since~$\TotK=0$, the first term at the r.h.s. of this identity
tends to zero as~$n\to\infty$.
The shifted quadratic form in the second term
can be written as a sum of three terms:
$$
  Q_1(\theta,\varphi_n\chi_1)
  =(\theta,2M_u\,\varphi_n\chi_1')
  +\left(\nabla\theta\chi_1,\nabla\varphi_n\right)
  -2\left(\theta\nabla\chi_1,\nabla\varphi_n\right),
$$
where the last two terms tend to zero as~$n\to\infty$ by means of
the Schwarz inequality, the estimates~(\ref{g<G<g}) and Lemma~\ref{Huber},
while an explicit calculation gives that the first integral
is equal to~$-(j,M\varphi_n)_g$ which tends to 
a \emph{non-zero} number~$-(j,M)_g$.
Since~$\theta$ does not depend on~$n$, one gets
$$
  Q_1[\varphi_n\chi_1+\varepsilon\theta]
  \,\xrightarrow[n\to\infty]{}\,
  -2\varepsilon\,(j,M)_g +\varepsilon^2 Q_1[\theta],
$$
which may be made negative by choosing~$\varepsilon$ sufficiently
small and of an appropriate sign.
\hfill\qed

\subsection{
Conditions (b) and (c)
}
Here we use the trial function $\psi_n(x,u):=(1+M(x)u)\,\varphi_n(x)\chi_1(u)$.
Since
\begin{eqnarray}\label{gradient}
  \nabla\psi_n(\cdot,u)
&=& (1+Mu) (\nabla\varphi_n) \chi_1(u)
  + (\nabla M) u \, \varphi_n \chi_1(u) \nonumber \\
&& + \, \big(
  (1+Mu) \kappa_1 \, \varphi_n \chi_1'(u) + M \varphi_n \chi_1(u)
  \big) \nabla u,
\end{eqnarray}
it is easy to see that~$\psi_n \in \Dom{Q}$ provided
$\nabla_{\!g}M \in \sii_\mathrm{loc}(\Sigma)$.
In this context and for further considerations,
we recall that the curvatures~$K$ and~$M$ are uniformly bounded,
\cf~$\Hii$. One has
\begin{eqnarray}\label{second}
  Q_1[\psi_n]
& \leq &
  2 \left(\left(1+a\|M\|_\infty \right)^2\| |\nabla\varphi_n|\,\chi_1 \|^2
  +a^2\| |\nabla M| \, \varphi_n\chi_1 \|^2\right) \nonumber \\
&& + \left(\varphi_n,(K-M^2)\varphi_n\right)_g
  +\frac{\pi^2-6}{12\kappa_1^2}
  \left(\varphi_n,K M^2\varphi_n\right)_g.
\end{eqnarray}
The inequality giving the factor~$2$ comes from the first line
at the r.h.s. of~(\ref{gradient}) and is established
by means of Minkovski's inequality and evident estimates.
The second line of~(\ref{second}) is the result of a direct calculation
and concerns the terms of the second line of~(\ref{gradient}).

We start by checking the sufficient condition~(c) of Theorem~\ref{thm.main}.
If~$\nabla_{\!g}M$ is $\sii$-integrable and~$\Hi$ holds true,
then all the terms at the r.h.s. of~(\ref{second}) tend to finite
values as~$n\to\infty$, except for the first integral at the second line
which tends to~$-\infty$ due to the assumption~$\TotM=+\infty$.
Hence we can find a finite~$n_0$ such that $Q_1[\psi_{n_0}]<0$.

There are two observations which lead to the condition~(b).
Firstly, the integral containing~$K-M^2$ in~(\ref{second})
is always negative for any non-planar and non-compact surface,
which can be seen by rewriting the difference of curvatures by means
of the principal curvatures, \ie~$K-M^2=-\frac{1}{4}(k_+-k_-)^2$.
Secondly, the first term at the r.h.s. of~(\ref{second}) tends
to zero as~$n\to\infty$ because of~(\ref{g<G<g}) and Lemma~\ref{Huber},
and the remaining ones vanish for~$n$ fixed as~$a \to 0$.
(For the latter we recall that~$\kappa_1^{-2}$ is proportional to~$a^2$.)
Hence we can find a sufficiently large~$n_0$ such that the sum
of the first term at the r.h.s. of~(\ref{second})
and the first integral at the second line of~(\ref{second}) is negative,
and then choose the layer half-width~$a$
so small that~$Q_1[\psi_{n_0}]<0$.
\hfill\qed

\subsection{
Condition (d)
}
Let~$\Sigma$ contain a cylindrically symmetric end~$E$
with a positive total Gauss curvature, $\TotK_E>0$.

Let us recall first the strategy employed in~\cite{DEK2}
to prove the existence of bound states in layers built
over surfaces of revolution diffeomorphic to~$\Real^2$
with a positive total Gauss curvature, \ie~$E=\Sigma$.
The essential ingredient is supplied by an information
about the behaviour of the mean curvature~$M$ at infinity.
In particular, if~$\TotK>0$, then~$|M|(\det{g})^\frac{1}{2}$ is bounded
but does not vanish at infinity of~$\Sigma$
and neither~$M$ nor~$M^2$ are integrable in~$\si(\Sigma)$.
On the other hand, the Gauss curvature is supposed to be integrable, \cf~$\Hi$.
Constructing an appropriate family of trial functions~$\{\Psi_n\}_{n\in\Nat}$
that is localised at infinity
(\ie,
$
  \forall\,\textrm{compact}\ \Omega_\mathrm{c}\subset\Omega \
  \exists n\in\Nat : \ \supp\Psi_n\cap\Omega_\mathrm{c}=\emptyset
$%
)
one succeeds to eliminate the contribution of the Gauss curvature
and, at the same time, to ensure that~$Q_1[\Psi_n]$
remains negative as~$n\to\infty$.
We refer to the proof of Theorem~6.1 in~\cite{DEK2} for more details
and an explicit form of~$\{\Psi_n\}_{n\in\Nat}$.

\begin{figure}[h]
\begin{center}
\epsfig{file=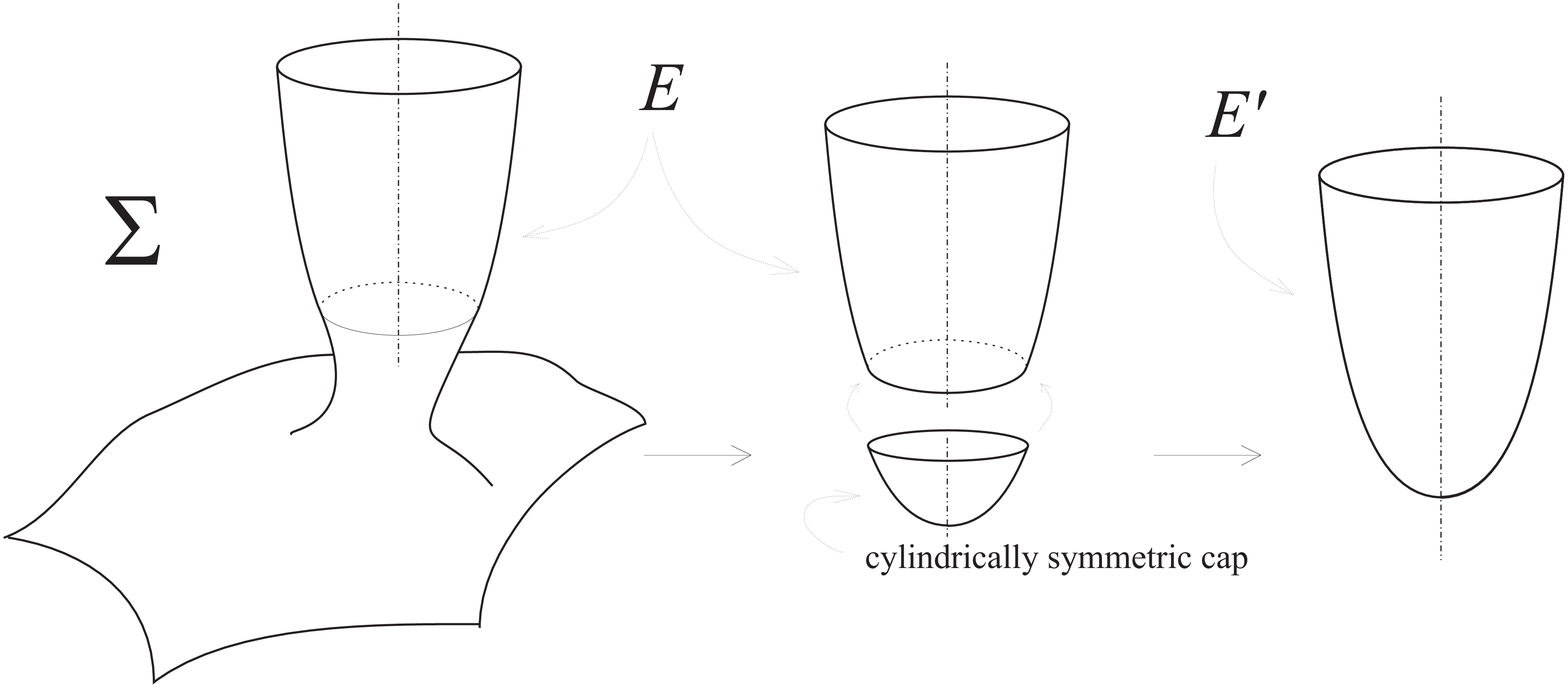,width=0.7\textwidth}
\end{center}
\caption{
Construction of a simply connected surface of revolution~$E'$
from a cylindrically symmetric end~$E \subset \Sigma$.
}
\label{Fig.symmetry}
\end{figure}

The fact that the family of trial functions was localised at infinity
makes it possible to extend the proof to our more general situation.
If~$E\not=\Sigma$, we construct from~$E$
a new cylindrically symmetric surface~$E'$
diffeomorphic to~$\Real^2$ by attaching smoothly to it
a cylindrically symmetric \emph{cap},
\ie, a simply connected surface with a compact boundary
(see Figure~\ref{Fig.symmetry}).
Since the attached surface is cylindrically symmetric
and simply connected, its total Gauss curvature cannot be negative,
which can be seen by the Gauss-Bonnet theorem
and a natural parameterisation, \cf~\cite[Sec.~6]{DEK2}.
Consequently, the total Gauss curvature of~$E'$
will not be less than the value~$\TotK_E$.
Since the latter is positive by assumption,
the mean curvature of~$E'$ behaves at infinity
like required for the use of~$\{\Psi_n\}_{n\in\Nat}$, which
proves the existence of spectrum below~$\kappa_1^2$
for the layer about~$E'$.
However, the identical asymptotic behaviour holds
for the mean curvature of~$E$ as well.
Hence, in order to establish the desired spectral result for the initial~$\Omega$,
it is sufficient to construct the sequence~$\{\Psi_n\}_{n\in\Nat}$
only at the infinity of the cylindrically symmetric layer
built over the end~$E$.
\hfill\qed
%
%

\section{Concluding Remarks}
The main interest of this paper was the Dirichlet Laplacian, $-\Delta_D^\Omega$,
in the layer region~$\Omega$ defined as a tubular neighbourhood
of a complete non-compact surface embedded in~$\Real^3$.
Using an intrinsic approach to the geometry of~$\Omega$,
the conditions of the original paper~\cite{DEK2},
sufficient to guarantee the existence of bound states below
the essential spectrum of~$-\Delta_D^\Omega$,
were significantly extended to layers built over general surfaces
without any strong topological restrictions;
see Theorem~\ref{thm.main} for the summary of the main results.

An important open problem is to decide whether the discrete
spectrum exists also for layers over surfaces with~$\TotK>0$
such that none of the conditions \mbox{(b)--(d)}
of Theorem~\ref{thm.main} is satisfied.
(We remind that, due to Corollary~\ref{Cor.CV},
it concerns surfaces diffeomorphic to~$\Real^2$ only.)
In view of the condition~(c),
it would be very desirable to prove the following conjecture
\begin{equation}\label{conjecture}
  \TotK>0 \ \Longrightarrow \ \TotM=+\infty.
\end{equation}
Taking into account the definition of~$K$ and~$M$ by means of
the principal curvatures, it may seem that there is no reason
to expect this property.
However, the principal curvatures cannot be regarded as arbitrary
functions because the first and second fundamental forms of~$\Sigma$
have to satisfy some integrability conditions
(the Gauss and Codazzi-Mainardi equations).
Note that we have proved the conjecture~(\ref{conjecture})
for cylindrically symmetric surfaces in~\cite{DEK2}.

Finally, interesting spectral results are expected
if the ambient space~$\Real^3$
is replaced by an Euclidean space of higher dimension
(more complicated normal bundle of~$\Sigma$)
or even by a general Riemannian manifold
(non-trivial structure of the ambient curvature tensor).

\section*{Acknowledgments}
The authors wish to thank Pierre Duclos for useful discussions.
The work has been partially supported by 
the ``ACI'' programme of the French Ministry of Research
and GA AS\,CR grant IAA 1048101.

%
%
\bibliography{bib}
\bibliographystyle{aip}
%
%
\end{document}